\documentclass[runningheads]{llncs}
\usepackage{graphicx}
\usepackage{booktabs}
\usepackage{multirow}
\usepackage{amssymb}
\usepackage{xcolor}
\begin{document}
%
\title{STRESS: Super-Resolution for Dynamic Fetal MRI using Self-Supervised Learning}
\titlerunning{STRESS}

\author{Junshen Xu\inst{1} \and
Esra Abaci Turk\inst{2} \and
P. Ellen Grant\inst{2,3} \and
Polina Golland\inst{1,4} \and
Elfar Adalsteinsson\inst{1,5}
}

%
\authorrunning{J. Xu et al.}
%
\institute{Department of Electrical Engineering and Computer Science, MIT, \\ Cambridge, MA, USA \\\email{junshen@mit.edu} \and
Fetal-Neonatal Neuroimaging and Developmental Science Center, \\ Boston Children’s Hospital, Boston, MA, USA \and
Harvard Medical School, Boston, MA, USA \and
Computer Science and Artificial Intelligence Laboratory, MIT, \\
Cambridge, MA, USA \and
Institute for Medical Engineering and Science, MIT, Cambridge, MA, USA}

\maketitle              
\setcounter{footnote}{0}

\begin{abstract}
Fetal motion is unpredictable and rapid on the scale of conventional MR scan times. Therefore, dynamic fetal MRI, which aims at capturing fetal motion and dynamics of fetal function, is limited to fast imaging techniques with compromises in image quality and resolution. Super-resolution for dynamic fetal MRI is still a challenge, especially when multi-oriented stacks of image slices for oversampling are not available and high temporal resolution for recording the dynamics of the fetus or placenta is desired. Further, fetal motion makes it difficult to acquire high-resolution images for supervised learning methods. To address this problem, in this work, we propose STRESS ({\bf S}patio-{\bf T}emporal {\bf R}esolution {\bf E}nhancement with {\bf S}imulated {\bf S}cans), a self-supervised super-resolution framework for dynamic fetal MRI with interleaved slice acquisitions. Our proposed method simulates an interleaved slice acquisition along the high-resolution axis on the originally acquired data to generate pairs of low- and high-resolution images. Then, it trains a super-resolution network by exploiting both spatial and temporal correlations in the MR time series, which is used to enhance the resolution of the original data. Evaluations on both simulated and \textit{in utero} data show that our proposed method outperforms other self-supervised super-resolution methods and improves image quality, which is beneficial to other downstream tasks and evaluations. 

\end{abstract}

\section{Introduction}

Fetal magnetic resonance imaging (MRI) is an important approach for studying the development of fetal brain \textit{in utero}~\cite{Saleem2014Fetal} and monitoring fetal function~\cite{luo2017vivo}. Due to unpredictable and rapid fetal motion, dynamic fetal MRI, which aims at capturing fetal motion and dynamics of fetal function, is limited to fast imaging techniques, such as single-shot Echo-planar imaging (EPI)~\cite{diogo2019echo}, with severe compromises in signal-to-noise ratio (SNR) and image resolution.

Super-resolution (SR) methods is frequently applied to fetal MRI to improve image quality. One well-established category of super-resolution methods for fetal MRI is based on slice-to-volume registration (SVR)~\cite{kuklisova2017distortion,uus2020deformable,gholipour2010robust}. In these methods, multiple stacks of slices at different orientations are acquired, which are then registered to reconstruct a static and motion-free volume of the chosen region of interest (ROI). However, multi-oriented stacks for oversampling the ROI may not available. Besides, in some applications, instead of a static ROI, a time series of MR volumes capturing the dynamics of fetal brain, body or placenta is of interest~\cite{kochunov2010fetal,xu2019fetal,luo2017vivo,turk2020placental}. For example, in~\cite{xu2019fetal} and ~\cite{luo2017vivo}, interleaved multi-slice EPI time series are used for fetal body pose tracking and placental function analysis respectively. Thus, it is a still a challenge to enhance the resolution in dynamic fetal MRI. 

Although supervised super-resolution methods achieved state-of-the-art results in natural images~\cite{lim2017enhanced,zhang2018residual}, the acquisition of HR MRI data with adequate SNR is time consuming and prone to motion artifacts, especially in fetal MRI. To avoid the need for HR data in supervised leanring, self-supervised super-resolution (SSR) methods have been developed, which utilize internal information from LR images for super-resolution. For instance, the ZSSR~\cite{shocher2018zero} method downsample the LR images to generate lower resolution (LR$_2$) images and train a network to learn a mapping from LR$_2$ to LR, which is then applied to the original LR images to estimate the HR images. Similar ideas are also explored in the field of MRI~\cite{jog2016self,zhao2020smore}. Zhao \textit{et al.} extended~\cite{jog2016self} and proposed SMORE~\cite{zhao2020smore} for SSR of MR volume with anisotropic resolution where the information along the LR axis are learned from the other two HR axes. They blur the volume along the one of the HR axes, extract pairs of training samples to train a network and use it to enhance resolution along the LR axis. However, these methods only applied to a single slice or a stack of images and cannot utilize the temporal information in dynamic imaging.

In this work, we propose a SSR framework for dynamic fetal MRI with interleaved acquisition, named STRESS ({\bf S}patio-{\bf T}emporal {\bf R}esolution {\bf E}nhancement with {\bf S}imulated {\bf S}cans). Using the characteristic of interleaved slice acquisition, we perform simulated acquisitions on the originally acquired data to generate pairs of low- and high-resolution images. We then train a SR network on the extracted data, which exploits both internal spatial information within each frame and temporal correlation between adjacent frames. A optional self-denoising network is also introduced to this framework, when input images are of low SNR. We evaluate the STRESS framework on both simulated and \textit{in utero} data to demonstrate that it can not only enhance resolution of dynamic fetal imaging but also improve performance of downstream tasks.

\section{Methods}

\begin{figure}[t]
\centering
\includegraphics[width=\textwidth]{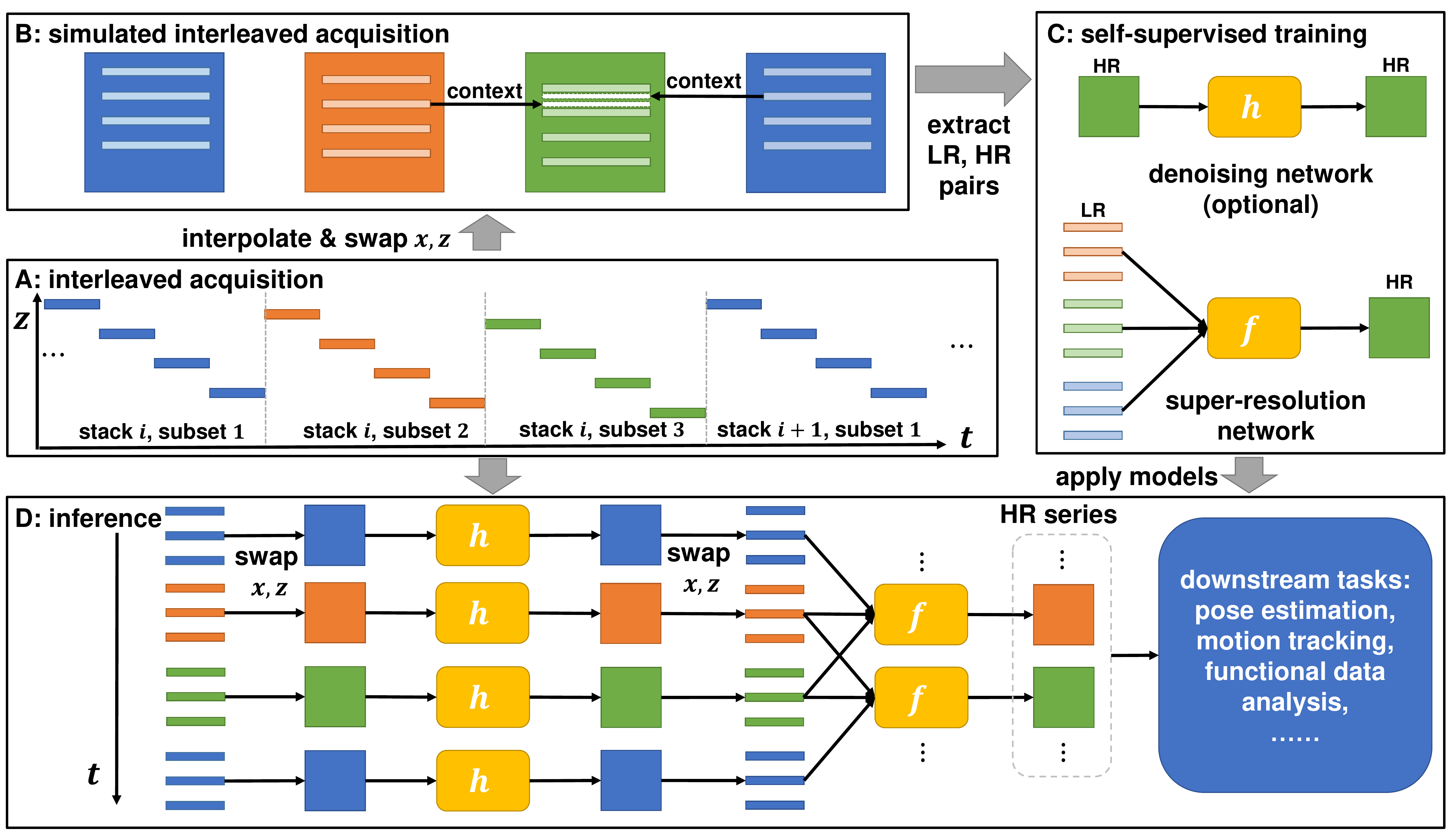}
\caption{The proposed STRESS workflow. A: Interleaved MRI acquisitions, e.g., $N_I=3$. B: Acquired MR data are binned into different time frames. These frames are interpolated and transposed to produce a simulated object with motion. Then, we simulate a interleaved MR scan on this object and extract low- and high-resolution pairs from them C: We train the denoising network (optional) and super-resolution network in self-supervised manners. D: We apply the trained models to the originally or newly acquired data to generate a high-resolution MR volume series, which can be further used for other downstream tasks.}
\label{fig:method}
\end{figure}

Fig. \ref{fig:method} shows the workflow of the proposed STRESS method, which can be divided into four parts: 1) interleaved slice acquisition, 2) simulated acquisition, 3) self-supervised training, and 4) inference. The details of each part are described in the following sections.

\subsection{Interleaved acquisition}

Interleaved slice acquisition is a widely used technique to avoid cross-excitation artifacts~\cite{dowling2009nonrigid}. The number of slices skipped between two consecutive slice acquisitions is often referred to as the interleave parameter~\cite{parker2014retrospective}, $N_I$. For example, when $N_I=2$, even slices are acquired after odd slices. Each image stack in interleaved acquisition are divided into $N_I$ interleaved subsets. In dynamic imaging, multiple stacks are acquired. For simplicity, we refer to the $i$-th subset in the $j$-th stack as time frame $F_k$, where the index $k=N_I\times(i-1)+j$. The acquisition time of each frame is only $1/N_I$ of the whole stack, making inter-slice motion artifacts within each frame milder. However, the spatial resolution of each frame along the interleaved axis is also reduced by a factor of $N_I$. Therefore the interleave parameter can be considered as a trade-off between between spatial and temporal resolutions. 

Our goal is to improve the spatial resolution of each frame to generate a HR MR series that has enough temporal resolution to capture fetal dynamics. Let $V_t(x,y,z)$ be the 3D dynamic object to be scanned, where $t$ is time and $(x,y,z)$ are the spatial variables. The acquisition of a slice at time $t$ and location $z$ is $V_t(\cdot,\cdot,z)$ 
Therefore, the $k$-th frame can be written as a set of slices, $F_k=\{V_t(\cdot,\cdot,z)|t=t(k,z), z\in\mathcal{Z}_k\}$, where $t(k,z)$ is the time when the slice at location $z$ of the $k$-th frame is acquired, and $\mathcal{Z}_k$ is the set of slice locations in the $k$-th frame.

\subsection{Simulated interleaved acquisition}

To generate HR and LR pairs for training a SSR network, we simulate the interleaved MR acquisition process with the acquired data. For each frame $F_k$, we interpolate it to make it an isotropic 3D volume denoted by $\widetilde{F}_{k}(x,y,z)$. Then we swap the $x$- and $z$- axis\footnote{We use $x$-axis here to keep the notation simple. In fact any axis within the $x$-$y$ plane can be used.} and result in a new 3D function $\widetilde{F}_{k}^T$, i.e., $\widetilde{F}_{k}^T(x,y,z)=\widetilde{F}_{k}(z,y,x)$. $\widetilde{F}_{k}^T(x,y,z)$ is an object of high resolution along the $z$-axis and having motion similar to $V_t$. Therefore, we can simulate interleaved acquisition along the $z$-axis to produce training pairs. The acquired frame in the simulated scan can be written as $S_{k} = \{\widetilde{F}_{k}^T(\cdot,\cdot,z)|z\in\mathcal{Z}_{k}\}$. Let $\widetilde{S}_{k}$ be the volume generated by interpolating $S_{k}$ along the $z$-axis. We can see that the $y$-$z$ planes of $\widetilde{S}_k$ and $\widetilde{F}_k$, i.e., $\widetilde{S}_{k+l}(x,\cdot, \cdot)$ and $\widetilde{F}_{k}^T(x,\cdot, \cdot)$ are pairs of LR and HR images. Besides, it is worth noting that the adjacent time frames provide contexts for estimating the missing slices in the target frame (Fig. \ref{fig:method} B). Therefore, it would be easier to learn a mapping from $\{\widetilde{S}_{k+l}(x,\cdot, \cdot)\}_{l=-L}^L$ to $\widetilde{F}_{k}^T(x,\cdot, \cdot)$, where $L$ is the number of time frames used from each side.

\subsection{Self-supervised training}

{\bf Super-resolution: } We extract image patches with size of $P\times P$ from the series of images, $\{\widetilde{S}_{k+l}(x,\cdot, \cdot)\}_{l=-L}^L$, and concatenate them along the channel dimension to form input tensors $I_{LR}\in\mathbb{R}^{P\times P\times (2L+1)}$. Patches at the same spatial locations are also extracted from $\widetilde{F}_{k}^T(x,\cdot, \cdot)$ as targets and denoted as $I_{HR}\in\mathbb{R}^{P\times P}$. A network $f$ is trained to learn the mapping between $I_{LR}$ and $I_{HR}$. L1 loss is used to improve the output sharpness, i.e., $\mathcal{L}=||f(I_{LR})-I_{HR}||_1$. We adopt the EDSR~\cite{lim2017enhanced} architecture for the SSR network $f$, with 16 residual blocks~\cite{he2016deep} and 64 feature channels.

{\bf Blind-spot denoising: }Many fast imaging techniques for capturing fetal dynamics, e.g., EPI, suffer from low SNR~\cite{gholipour2014fetal}. Applying super-resolution algorithms to noisy images tends to emphasize image noise and results in images of low quality. To address this problem, we introduce an optional denoising network $h$ to our framework, which can be apply when the original acquired images are of low SNR. The network $h$ is a blind-spot denoising network (BDN)~\cite{laine2019high}, i.e., the receptive field of $h$ doesn't contain the central pixel. Therefore, when we train the network $h$ to recover the input image $I$ by minimizing the mean squared error, $||h(I)-I||_2^2$, the network will not become the identity function. Instead, $h(I)$ will approximate the mean of $I$, so that $h(I)$ can be considered as the denoised image. If BDN is enabled, we first train the denoising network $h$ with images $I=\widetilde{F}_{k}^T(x,\cdot, \cdot)$. Then, when training the SSR network $f$, we replace the target $I_{HR}$ with $h(I_{HR})$ and the loss becomes $\mathcal{L}=||f(I_{LR})-h(I_{HR})||_1$. 

{\bf Training details: }We set $L=N_I/2$ and $P=64$, if not specifically indicated. All neural networks are trained on a 
Nvidia Tesla V100 GPU using an Adam optimizer~\cite{kingma2017adam} with a learning rate of $1\times 10^{-4}$ for 30000 iterations. We use batch sizes of 64 and 16 for network $f$ and $h$ respectively, which depend on GPU memory. Training images are randomly flipped along the two axes for data augmentation. Our models are implemented with PyTorch 1.5~\cite{paszke2017automatic}.

\subsection{Inference}

After training the models, we can apply them to the original or newly acquired data. If BDN is enabled, we first perform image denoising on each frame by applying $h$ to each slice, such that $F_k$ becomes $\{h(V_t(\cdot,\cdot,z))|t=t(k,z), z\in\mathcal{Z}_k\}$. Then, we interpolate it to generate a volume, $\widetilde{F}_{k}(x,y,z)$. Finally, the trained super-resolution network $f$ is applied to the $y$-$z$ plane of $\widetilde{F}_{k}(x,y,z)$ and its neighboring frames, which yields a super-resolved estimate $\hat{V}_k$, i.e., $\hat{V}_k(x,\cdot, \cdot) = f(\{\widetilde{F}_{k+l}(x,\cdot, \cdot)\}_{l=-L}^L)$. This process is repeated for all $k$ until we get a HR estimation of the whole series, which can be used for other downstream tasks.

\section{Experiments and Results}

In the experiments, we apply the following methods to fetal MR volume series: 1) cubic B-spline interpolation along the interleaved axis; 2) interpolation along the temporal direction (TI); 3) spatio-temporal interpolation (STI); 4) SMORE~\cite{zhao2020smore} and 5) STRESS. In SMORE, we adopt the same super-resolution network architecture and the same training hyperparameters as STRESS for fair comparison. The reference PyTorch implementation for STRESS is available on GitHub\footnote{\url{https://github.com/daviddmc/STRESS}}

\subsection{CRL fetal dataset}
The CRL fetal atlas~\cite{gholipour2017normative} consist of T2-weighted fetal brain MRI with gestational age (GA) ranging from 21 to 38 weeks. The images are reconstructed to volume with size of $135\times189\times155$ and isotropic resolution of 1 mm. To simulate fetal motion, we use the fetal landmark time series in~\cite{xu2019fetal}. Specifically, we use two eyes and the midpoint of two shoulder to define the fetal pose and apply affine transformation to the MR volume to generate motion trajectories. There are 77 time series with length from 20 to 30 minutes in the landmark dataset. We randomly sample 10 1-min intervals from each series then apply the motion to the volumes, resulting in $18\times77\times10=13860$ data. We use 70\% data for training and validation, 30\% for test, data in the test set have different GAs from training and validation sets. We simulate MR scans with $N_I=2,4$ and 6, in-plane resolution of $1 \mbox{mm}\times1 \mbox{mm}$ and slice thickness of $1\mbox{mm}$. SR methods are applied to the noise-free data and also noisy data corrupted by Rician noise~\cite{gudbjartsson1995rician} with standard deviation $\sigma = 3\%$ of the maximum intensity. BDN is enabled when there is noise.

\begin{figure}[h]
\centering
\includegraphics[width=\textwidth]{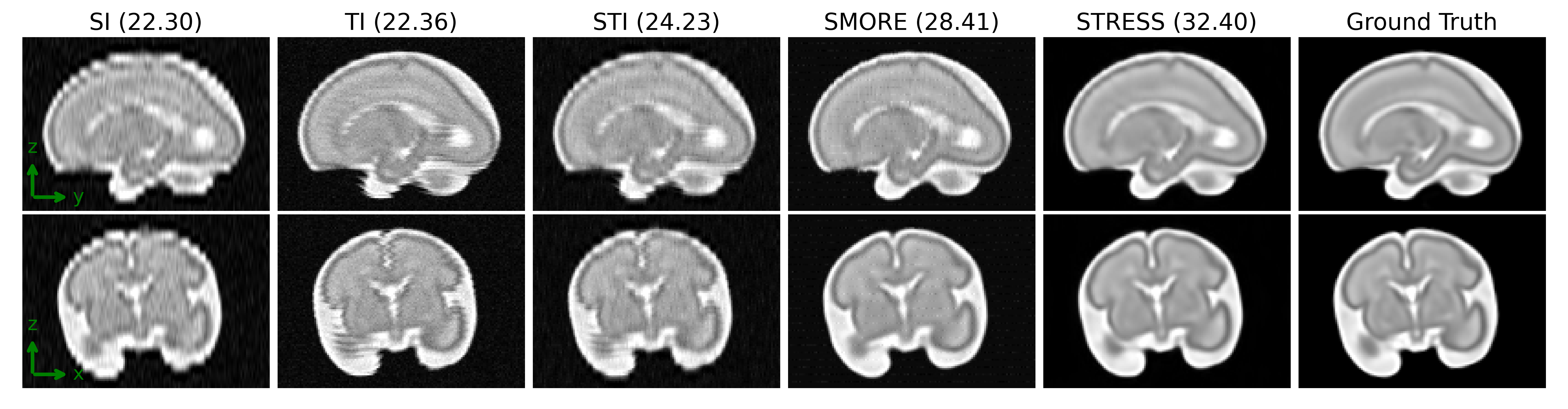}
\caption{Visual results from CRL fetal dataset ($N_I=4$, Rician noise $\sigma=3\%$ of the maximum intensity), numbers in the parentheses are PNSR with ground truth data as reference.}
\label{fig:sim_vis}
\end{figure}

Table~\ref{tab:sim} shows the peak signal-to-noise ratio (PSNR) and structural similarity index (SSIM)~\cite{wang2004image} comparing to the ground truth. PSNR and SSIM are computed within a mask of non-background voxels. The proposed STRESS method outperforms the competing methods at different interleave parameters, with and without noise. Fig.~\ref{fig:sim_vis} shows example slices of super-resolution results with $N_I=4$ and Rician noise. Visual results also indicates that the outputs of STRESS have better image quality.

\begin{table}[h]
\centering
\begin{tabular}{c|cc|cc|cc|cc|cc|cc}
\toprule
\multirow{3}{*}{Models} & \multicolumn{4}{c|}{$N_I=2$} & \multicolumn{4}{c|}{$N_I=4$} & \multicolumn{4}{c}{$N_I=6$}\\
\cline{2-13}
 & \multicolumn{2}{c|}{w/o noise} & \multicolumn{2}{c|}{w/ noise} &  \multicolumn{2}{c|}{w/o noise} & \multicolumn{2}{c|}{w/ noise}&  \multicolumn{2}{c|}{w/o noise} &  \multicolumn{2}{c}{w/ noise} \\
\cline{2-13}
   & PSNR & SSIM & PSNR & SSIM & PSNR & SSIM & PSNR & SSIM & PSNR & SSIM & PSNR & SSIM\\
\midrule
SI     & 32.69 & .9883 & 28.42 & .8849 & 23.90 & .9049 & 22.98 & .8114 & 19.71 & .7422 & 19.39 & .6686 \\
TI  & 29.01 & .9111 & 25.31 & .8258 & 29.21 & .9076 & 25.48 & .8273 & 28.60 & .9084 & 25.52 & .8288 \\
STI   &31.29 &  .9682 & 27.94 & .8846 & 26.87 & .9390 & 25.75 & .8711 & 23.89 & .8769 & 23.37 & .8182 \\
SMORE     & 36.19 & .9895 & 30.38 & .9006 & 31.36 & .9687 & 28.57 & .8916 & 25.29 & .8703 & 24.27 & .8093 \\
STRESS    & \underline{36.77} & \underline{.9921} & \underline{33.51} & \underline{.9702} & \underline{34.56} & \underline{.9873} & \underline{32.81} & \underline{.9655} & \underline{28.98} & \underline{.9480} & \underline{28.24}  & \underline{.9213} \\
\bottomrule
\end{tabular}
\caption{PSNR and SSIM of the super-resolution results on the CRL dataset, where 'w/ noise' means adding Rician noise with $\sigma=3\%$ of the maximum intensity. The best results are underlined.}
\label{tab:sim}
\end{table}

In addition, we also evaluate the performance of the STRESS method with and without BDN under different noise levels ($\sigma=1\%, 3\%$, and $5\%$ of the maximum intensity). The results are shown in Table~\ref{tab:noise}. We can observe that the BDN makes a larger contribution to the performance of STRESS as the noise level increases.

\begin{table}[h]
\centering
\begin{tabular}{c|cc|cc|cc}
\toprule
\multirow{2}{*}{Models} & \multicolumn{2}{c|}{$\sigma=1\%$} & \multicolumn{2}{c|}{$\sigma=3\%$} & \multicolumn{2}{c}{$\sigma=5\%$}\\
\cline{2-7}
   & PSNR & SSIM & PSNR & SSIM & PSNR & SSIM \\
\midrule
STRESS w/o BDN & 33.96 & .9764 & 30.69 & .9219 & 28.29 & .8559 \\
STRESS w/ BDN & 33.99 & .9826 & 32.81 & .9655 & 31.09 & .9425 \\
\bottomrule
\end{tabular}
\caption{Evaluations of STRESS with and without BDN under different noise levels ($N_I=4$).}
\label{tab:noise}
\end{table}

\subsection{Fetal EPI dataset}
We also evaluate our method with an \textit{in utero} fetal EPI dataset in~\cite{luo2017vivo}, which consist of 111 volumetric MRI time series at a gestational age ranging from 25 to 35 weeks. MRIs were acquired on a 3T Skyra scanner (Siemens Healthcare, Erlangen, Germany). Interleaved, multislice, single-shot, gradient echo EPI sequence was used for acquisitions with in-plane resolution of $3\mbox{mm}\times3\mbox{mm}$, slice thickness of 3 mm, average matrix size of $120\times120\times80$; TR=$5-8$s, TE=$32-38$ms, FA=90$^{\circ}$, $N_I=2$. Each subject was scanned for 10 to 30 min. We remove half of the slices at each frame to generate data with $N_I=4$. We use 92 EPI series for training and 19 for testing. Due to the large voxel size in acquisition and the relatively high SNR, we disable BDN on this dataset. Besides, some volumes have matrix size less than 64, so we use $P=32$ in this experiment. 

Since ground truth is not available for the \textit{in utero} dataset, we use the removed slices as reference to compute PSNR and SSIM. To further evaluate the quality of output images, we use fetal keypoint detection as a downstream task, where 15 fetal keypoints (ankles, knees, hips, bladder, shoulders, elbows, wrists and eyes) are detected from each time frame. Ground truth labels are manually annotated on the original data with $N_I=2$. We apply a pretrained keypoint detection model~\cite{xu2019fetal} to the output volumes of each SR method. The percentage of correct keypoint (PCK)~\cite{andriluka20142d} are computed. $\mbox{PCK}(s)=N(s)/N\times100\%$, where $N$ is the total number of keypoints and $N(s)$ is the number of predicted keypoints with error less than threshold $s$.

\begin{figure}[h]
\centering
\includegraphics[width=\textwidth]{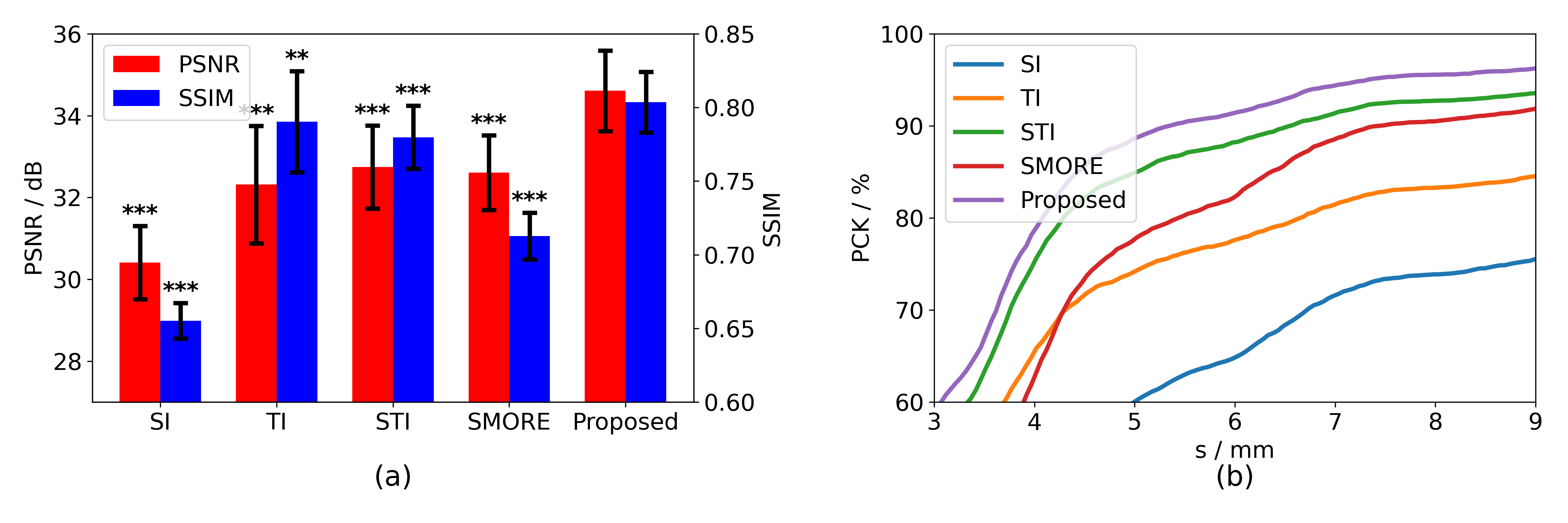}
\caption{Evaluation of super-resolution results from fetal EPI data with $N_I=4$. Left: PSNR and SSIM comparing to the reference in the $N_I=2$ data. Error bars show the corresponding standard deviations. $**$: p-value $<10^{-2}$, $***$: p-value $<10^{-3}$. Right: PCK curves for fetal landmark detection using a pretrained model.}
\label{fig:epi}
\end{figure}

Fig.~\ref{fig:epi} shows the evaluation of super-resolution results on the fetal EPI dataset. The proposed STRESS method achieves the highest PSNR and SSIM among all competing methods, which is also shown by the t-test. Besides, when using the super-resolution results for fetal keypoint detection, the results of STRESS also have the best performance in terms of PCK, indicating that the STRESS method is able to generate MR time series with high image quality which is beneficial to downstream tasks.
 
\begin{figure}[h]
\centering
\includegraphics[width=\textwidth]{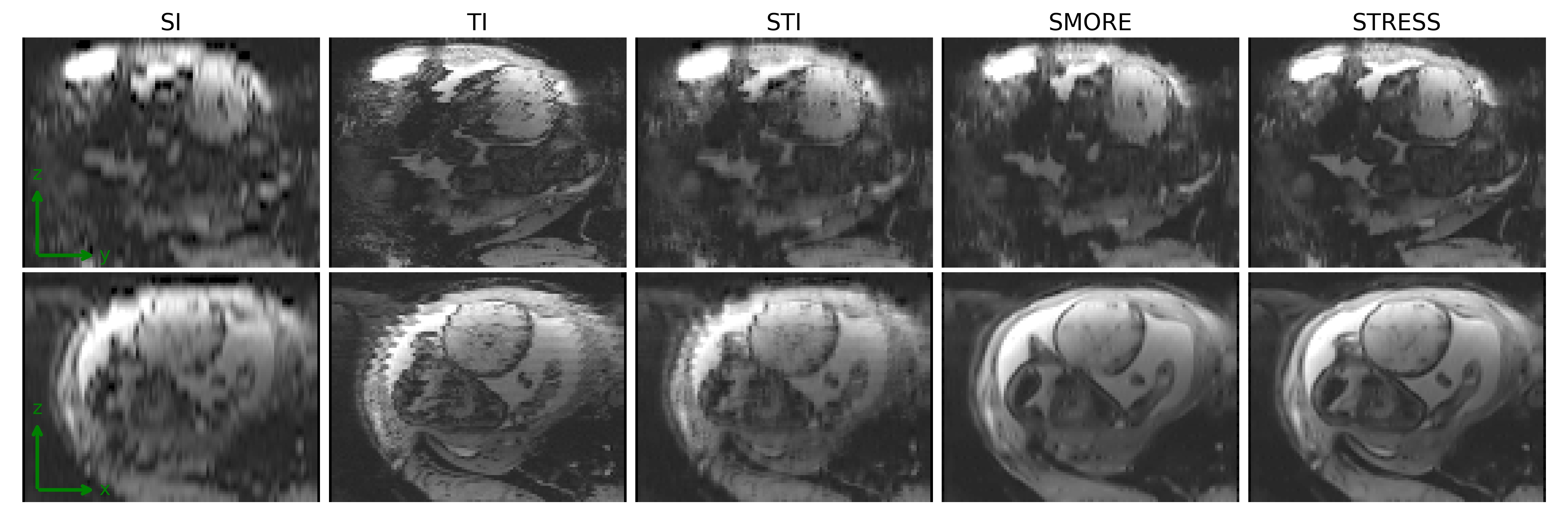}
\caption{Visual results from \textit{in utero} fetal EPI dataset.}
\label{fig:epi_vis}
\end{figure}

Fig.~\ref{fig:epi_vis} shows example slices of super-resolution results in one frame of the fetal MR series. We can see that the results of the proposed STRESS method have the best perceptual quality. The output of SI is very blurred, since it only interpolates along the $z$-axis. The TI and STI methods utilize temporal information with simple interpolation and therefore introduce severe inter-slice misalignment to the images. Although SMORE achieves better image quality than interpolation methods, the boundary of fetal brain is unclear in the outputs of SMORE. The reason is that SMORE only take a single frame as input without the temporal context, so that it cannot restore the details in the body parts that are corrupted by fetal motion, such as the fetal brain. STRESS, however, utilizes both spatial and temporal information of the scan data during the self-supervised training process, and therefore recovers more image details.

\section{Conclusions}

This paper presents STRESS, a self-supervised super-resolution framework for dynamic fetal imaging with interleaved slice acquisition. STRESS trains a SR network in a self-supervised manner, where low- and high-resolution training samples are extracted from simulated interleaved acquisitions. The SR network utilizes both internal spatial information within each frame and temporal correlation between adjacent frames to improve image quality and restore details corrupted by fetal motion. Evaluations on both simulated and \textit{in utero} data shows that STRESS outperforms other competing methods. The experiments also demonstrate that STRESS is beneficial when serving as a data pre-processing step for further downstream analysis.

\section*{Acknowledgements}
This research was supported by NIH U01HD087211, NIH R01EB01733 and NIH NIBIB NAC P41EB015902.

%
%
%
\bibliographystyle{splncs04}
\bibliography{ref.bib}
%




\end{document}